\documentclass{elsart}
\usepackage{graphicx}
\begin{document}
\begin{frontmatter}
\title{Density functional theory study of (OCS)$_2^-$}
\author{G. Bilalbegovi{\'c}}
\date{Chem. Phys. Lett. (2007)}
\address{Department of Physics, Faculty of Science, University of Zagreb, Bijeni{\v c}ka 32, 10000 Zagreb, Croatia}
\begin{abstract}

The structural and electronic  properties of the carbonyl sulfide dimer anion are calculated
using density functional theory within a pseudopotential method.
Three geometries are optimized and investigated:
C$_{2v}$ and C$_2$ symmetric,  as well as one asymmetric structure.
A distribution of an excess charge in three isomers are studied by the Hirshfeld method.
In an asymmetric (OCS)$_2^-$ isomer the charge is
not equally divided between the two moieties, but it is
distributed as OCS$^{-0.6} \cdot$ OCS$^{-0.4}$.
Low-lying excitation levels of three isomers are compared using the time-dependent density functional theory in the Casida approach.

\end{abstract}
\end{frontmatter}
\section{Introduction}
\label{intro}

Ionic nanoparticles of various sizes offer possibilities to investigate the change of a chemical bonding from molecules and clusters to the bulk.
Studies of ionic clusters are important for understanding of microscopic aspects of chemical reactions in
controlled environments \cite{Lineberger}.
Carbonyl sulfide may play an important role in an explanation of the origin of life on Earth. It acts as
a condensing agent in the formation of peptides from amino acids \cite{Leman}.
This material is known as a component of volcanic gases and it is present in the atmosphere of Earth
where it is a part of the global sulfur cycle. Carbonyl sulfide is also present in the atmosphere of Venus
and in many other astrophysical objects.
Therefore, OCS and related ions and clusters are important for various chemical and physical processes.
It has been found that
OCS$^{-}$ anion is metastable, but that an addition of OCS or H$_2$O molecules produces a stable structure \cite{Surber,Ananthavel}.

For homogeneous anionic clusters, such as (OCS)$_n^{-}$, it is important to
find out whether an excess electron is located in a vicinity of a
particular monomer, or it is shared between two or more fragments.
The photodissociation dynamics of (OCS)$_2^{-}$ has
been studied in the ion beam apparatus using a tandem time-of-flight
mass spectrometry \cite{Sanov}. The photoelectron imaging method has
been also applied in the studies of (OCS)$_2^-$
\cite{Surber,Sanov2}. The coexistence of two covalent OCS$_2^-$
isomers (having C$_{2v}$ and C$_2$ symmetry), as well as
electrostatically bound OCS$^{-} \cdot $ OCS have been proposed
\cite{Lineberger,Sanov}. The formation of (OCS)$_n^-$ cluster ions
have been studied using low-energy electron attachment to molecular
clusters (OCS)$_n$ seeded in helium carrier gas \cite{Barsotti}.
This experiment have been supported by the coupled cluster theory
calculations of electron affinity and vertical detachment energy of
the carbonyl sulfide \cite{Barsotti}.

Advanced experimental methods, such as photoelectron spectroscopy, are able to explore  various nanostructures. It is important to compare these results with different theoretical simulations of the same systems.
The properties of (OCS)$_2^{-}$ cluster have been studied using the Gaussian quantum chemistry code
in connection with the photodissociation experiments \cite{Sanov}.
A related (OCS)$_2^{+}$ complex has also been investigated using the Gaussian program \cite{McKee}.
The pseudopotential based density functional theory is
characterized by the computational efficiency and recent intensive
developments. It is nowadays possible to use this method for big
systems which exhibit complex chemical and physical processes. It
allows computations of larger systems than traditional quantum
chemistry methods. The pseudopotential time-dependent density
functional theory is promising and still in development for
many applications, for example in studies of charge transfer
processes. It is important to use density functional theory and
pseudopotentials to simulate some of the structures and processes
studied in experiments on time-resolved dynamics and solvation
in big anion clusters \cite{Lineberger}.
However, methods initially developed in
theoretical chemistry sometimes better describe dynamics of
electrons involved in chemical processes. Therefore, it is important
to test the pseudopotential density functional methods in studies on
small charged clusters. In the previous computational study of $(OCS)_2^-$, which has been done using the Gaussian package \cite{Lineberger,Sanov},
the charge population has not been analyzed, and a direct calculation of low-lying excited energy levels
of different isomers has not been presented.
In this work (OCS)$_2^{-}$ is studied using density functional theory (DFT) electronic structure calculations.
Three geometries of the carbonyl sulfide dimer anion are minimized and their structures, energies, the ground state electronic properties, as well as distributions of the lowest excitation levels are investigated.

\section{Computational Method}
\label{sec:2}

DFT based calculations have been recently used as a powerful
method for studies of bulk, surfaces, and nanostructures of various
materials \cite{Martin}. The structural and electronic properties of
(OCS)$_2^{-}$ are studied in this work using the plane wave density
functional Abinit code \cite{Abinit}. 
This code is robust, accurate and widely used. It has been applied by many users on various geometries and processes, and it is therefore well tested.
The local density
approximation of the exchange functional in the parametrization of
Perdew and Wang is  applied \cite{Perdew}.
The  (OCS)$_2^{-}$ ions (and related structures used for testing and
comparison) are positioned in the middle of the periodically
replicated simulation box. The side of the box for most
simulations was $30$ a.u.
The Brillouin zone is sampled
using the $\Gamma$ point. Relaxation of ions is carried out by
performing a series of self-consistent calculations. The
Broyden-Fletcher-Goldfarb-Shanno minimization method is used for a
structural optimization and all atoms are allowed to move.

In the Abinit package charged systems are
immersed in a neutralizing jellium background.  For such charged
particles artificial electrostatic interactions between the initial
system and its periodic images in the supercells are possible
\cite{Joannopoulos,Makov,Schultz}. However, the supercells used here
are sufficiently large to avoid these effects. To check this point several
calculations for larger cells ($35$ and $40$ a.u.) are done and the differences in
the total energy, bond lengths and angles are found to be $0.1 \%$  at the most.
In addition, the charge density plots do not change when the side
of the cell increases.

Because of a fragility of some possible (OCS)$_2^{-}$ structures, two kinds of pseudopotentials are used and the results
are compared.
First, the calculations are carried out using the Troullier-Martins (TM) pseudopotentials \cite{Troullier}, prepared
by the Fritz Haber Institute code \cite{Fuchs} and taken from the Abinit web page \cite{Abinit}. The cutoff of $35$ Ha is used for the TM pseudopotentials.
Calculations are also done using  the relativistic separable pseudopotentials in the form of Hartwigsen, Goedecker and Hutter (HGH) \cite{Hartwigsen}. The cutoff of 50 Ha is applied for the HGH pseudopotentials.
It has been found that the properties of CS, CS$_2$ and several oxide molecules
calculated using these pseudopotentials agree with experiments \cite{Hartwigsen}.
It is calculated in this work that both kinds of pseudopotentials (HGH and TM) produce a good agreement with experiments for the properties of OCS. This molecule
(in experiments, as well as in calculations described here)
stabilizes into a linear structure.
The S-C distance is 1.548 \AA{}
(the experimental value is 1.561 \AA{} \cite{Ananthavel}),
and the C-O distance is 1.160 \AA{} (the experimental result is 1.156 \AA{}) in the calculations using
the HGH pseudopotentials.
The calculated distances are 1.548 \AA{} and 1.148 \AA{}, respectively for the S-C and C-O bonds,
when the TM pseudopotentials are applied.
For the (OCS)$_2^{-}$ isomers,
only small numerical differences are found between
calculations carried out using the TM and HGH pseudopotentials.
It is important to report the result for an electron
affinity of the OCS molecule. However, it is not possible to stabilize (OCS)$^-$ structure
within the method used in this work  and this
is a necessary step for the calculation of electron affinity. The
absence of a stable (OCS)$^-$ anion in this calculation is in
agreement with the experimental results. This anion is not formed in
a standard ion source and almost no (OCS)$^-$ is detected in the
mass-spectrum \cite{Lineberger,Surber,Ananthavel,Sanov,Sanov2}.

The excitation energies are calculated within the time-dependent local density functional theory using the Casida
approach \cite{Gross,Casida} implemented in the Abinit code.
In this method the time-dependent density functional theory (TDDFT) equations are studied in the frequency domain.
The eigenvalue problem is solved for the matrix whose main part is formed of the squares of differences
between occupied and unoccupied Kohn-Sham electronic energies. The coupling matrix which includes the contribution
of the Coulomb and exchange-correlation interactions is also added to the Casida matrix.
It is known that low-lying excited states of small, chemically simple clusters and molecules are well represented
in the Casida approach. New algorithms in TDDFT are necessary for periodic systems and high-lying excited levels. In this work the Casida approach to TDDFT is used to compare low-lying excitation
energy levels of three (OCS)$_2^{-}$ isomers.

\section{Results and discussion}
\label{sec:3}

\begin{figure}
\begin{center}
\caption{The density functional optimized structures and distances of symmetric (OCS)$_2^{-}$: (a) C$_{2v}$,
(b) C$_2$. Bond lengths (in {\AA}ngstr{\" o}ms) and angles (in degrees) are shown.}
\label{fig:fig1}
\end{center}
\end{figure}

Three equilibrium structures of $(OCS)_2^-$ are stabilized within
approximations of density functional theory used in this work and
for both kinds of pseudopotentials. These calculations started from
initial geometries close to those proposed by Sanov
and coworkers \cite{Lineberger,Sanov}.
Several other initial structures
of (OCS)$_2^{-}$ were also considered, but none of them is found to
be stable. The equilibrium structures of (OCS)$_2^{-}$, as well as
corresponding bond lengths and angles calculated using the TM
pseudopotentials are shown in Figs. 1 and 2. In the asymmetric
(OCS)$_2^{-}$ structure (shown in Fig. 2) both OCS fragments  are
nonlinear. This also applies for the same isomer optimized using the
HGH pseudopotentials. The electron charge isosurface plot for the
asymmetric (OCS)$_2^{-}$ structure is also shown in Fig. 2.
Two moieties are bonded between the carbon atom of one fragment and the
sulphur atom of the other. The distance between these two atoms is 2.49 \AA.
Figures 2 and
3(c) show that a week covalent interaction between the two moieties
exists. In the structures found by Sanov and coworkers one 
isomer is asymmetric. This is an electrostatically bound structure
where one of the moieties is linear, and the charges are distributed
as OCS$^{-} \cdot $ OCS \cite{Lineberger,Sanov}.
Both asymmetric isomers, weakly covalent and
electrostatically bonded one, may form under experimental
conditions.
The binding energies are presented in
Table 1. The C$_{2v}$ structure is the most stable. The differences
in energy of about $\Delta E_1 = 0.25$ eV and $\Delta E_2 = 0.61$ eV
exist between the C$_{2v}$ structure and two other isomers. These results
also show that the binding energy of the most stable structure of
the C$_{2v}$ (OCS)$_2^{-}$ isomer is $\sim 3$ eV lower than one of
the neutral C$_{2v}$ (OCS)$_2$ cluster optimized by the same method.

\begin{figure}
\begin{center}
\caption{The density functional optimized asymmetric (OCS)$_2^{-}$
structure. The electron charge density, as well as bond lengths (in Angstr{\"o}ms) and angles (in degrees) are shown. The charge contour corresponds to a constant density equal to $0.02$ eV/a$_0^3$.}
\label{fig:fig2}
\end{center}
\end{figure}

\begin{table}
\caption{\label{tab:table1}
Binding energies (in eV) of three optimized anions of (OCS)$_2^{-}$ and a neutral  (OCS)$_2$ cluster
(having C$_{2v}$ symmetry)
calculated using the TM and HGH pseudopotentials.}
\begin{tabular}{ccccc}
\hline
Structure & C$_{2v}$ (OCS)$_2^{-}$  & C$_2$  (OCS)$_2^{-}$  & asymmetrical (OCS)$_2^{-}$ & (OCS)$_2$\\
\hline
TM  & -46.794 & -46.544 & -46.188 & -43.750    \\
HGH & -46.602 & -46.357 & -46.002 & -43.293   \\
\hline
\end{tabular}
\end{table}

The Hirshfeld method is a suitable technique to analyze the charge
density distribution in molecules decomposed into atomic fragments
\cite{Hirshfeld,Nalewajski}. In this method a molecule is divided
into atomic contributions proportional to the weight of the free
atom charge density. The free atoms situated in their corresponding
positions in the molecule define the promolecule. This reference
state of the promolecule is compared to the actual charge on the
real molecule induced by the formation of chemical bonds. The
Hirshfeld charges of (OCS)$_2^-$ are shown in Tables 2 and 3. The
third structure (the results shown in Table 3) is not symmetric and
charges are different for two atoms in three pairs of the same kind
(S, C, and O). A small positive charge is located on the carbon
atoms in C$_{2v}$ and C$_2$ symmetric structures, whereas a negative
charge is distributed on all atoms in asymmetric (OCS)$_2^{-}$. The
results for the neutral (OCS)$_2$ dimer (also presented in Table 2)
show that charges on all atoms become more negative in (OCS)$_2^-$.
The oxygen Hirshfeld charges in (OCS)$_2^-$ isomers are
substantially lower than in the water molecule and in the carbonyl
oxygen of the acetamide, but it is also higher than in the carbonyl
oxygen of the formaldehyde \cite{DeProft}. These Hirshfeld charges
for various molecules are calculated within different {\it ab
initio} computational programs. However, previous calculations (for
a different material) have been shown that the values of Hirshfeld
charges are not sensitive to the calculation technique
\cite{Goranka}. Table 3 shows that the excess charge in an asymmetric
(OCS)$_2^{-}$ structure is distributed as $-0.6:-0.4$. More bended
fragment of this isomer takes more, but not all of the negative
charge. The part of the excess charge is located on the other moiety
which is also bended.

\begin{table}
\caption{Hirshfeld charges of (OCS)$_2^{-}$ for two symmetric structures calculated using
the TM pseudopotentials. The charges on atoms in the C$_{2v}$ symmetric  neutral (OCS)$_2$ cluster are also shown.
The values for the HGH pseudopotentials are in the parenthesis.}
\label{table2}
\begin{tabular}{cccc}
\hline
Structure & C$_{2v}$ (OCS)$_2^{-}$ & C$_2$ (OCS)$_2^{-}$ & C$_{2v}$ (OCS)$_2$ \\
\hline
$\delta$Q(S) & -0.251 (-0.250)& -0.296 (-0.294)&  +0.049 (+0.049)\\
$\delta$Q(C) & +0.026 (+0.018)& +0.038 (+0.030)&  +0.100 (+0.092)\\
$\delta$Q(O) & -0.274 (-0.268)& -0.241 (-0.235)&  -0.149 (-0.141)\\
\hline
\end{tabular}
\end{table}

\begin{table}
\caption{Hirshfeld charges for the asymmetric (OCS)$_2^{-}$
structure calculated using the TM pseudopotentials (the values for
the HGH pseudopotentials are in the parenthesis). Atoms labeled with
``1'' belong to the moiety on the right in Fig. 2.} \label{table3}
\begin{tabular}{ccc}
\hline
Structure & Atom 1& Atom 2 \\
\hline
$\delta$Q(S) &-0.156 (-0.154) & -0.344 (-0.343) \\
$\delta$Q(C) &-0.006 (-0.016) & -0.010 (-0.016) \\
$\delta$Q(O) &-0.232 (-0.226) & -0.245 (-0.238) \\
\hline
\end{tabular}
\end{table}

\begin{figure}
\begin{center}
\caption{The HOMO states of (OCS)$_2^{-}$ isomers: (a) C$_{2v}$ symmetric, (b) C$_2$ symmetric, (c) asymmetric structure. 
The light (dark) isosurface corresponds to the negative (positive) part of the wave functions.}
\label{fig:fig3}
\end{center}
\end{figure}

Figure 3 presents the HOMO wave functions,
whereas Fig. 4 shows the energy levels 
of optimized
(OCS)$_2^{-}$ isomers.
Delocalized negative orbitals exist only in the C$_2$ symmetric (Fig. 3 (b)) and
asymmetric structure (Fig. 3(c)).
The $\sigma$-like bonding orbital exists between the carbon atom on the left fragment
and the sulphur atom on the right one in the asymmetric structure (Fig. 3(c)).
The details of electronic structure, such as
a distribution of levels and shapes of HOMO states, are different
for these three isomers. As in many other systems, these results
show that the chemical and physical properties of nanostructures are
determined not only by their size and the atoms they consist of. The
details of geometrical and electronic structure are also important.

\begin{figure}
\begin{center}
\includegraphics*[scale=0.6]{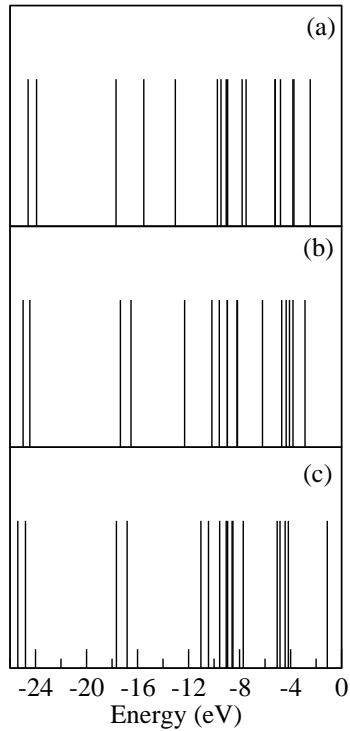}
\caption{Electron eigenvalues for (OCS)$_2^{-}$ isomers: (a) C$_{2v}$ symmetric, (b) C$_2$ symmetric, (c) asymmetric structure. Several levels are degenerate and almost degenerate, and therefore they are not distinguishable in the figure.}
\label{fig:fig4}
\end{center}
\end{figure}

\begin{figure}
\begin{center}
\includegraphics*[scale=0.6]{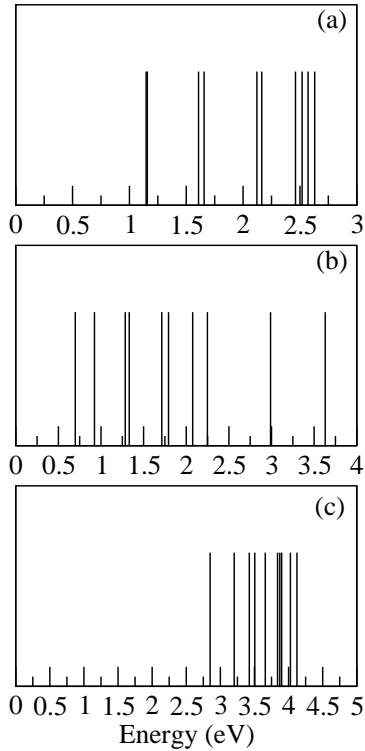}
\caption{The lowest excitation energies from the ground-state of (OCS)$_2^{-}$ isomers: (a) C$_{2v}$ symmetric, (b) C$_2$ symmetric, (c) asymmetric structure. The first ten mixed singlet and triplet energy levels are shown
(several levels are not distinguishable on these figures).}
\label{fig:fig5}
\end{center}
\end{figure}

In experimental studies of (OCS)$_2^{-}$  \cite{Surber,Sanov,Sanov2} it has been found that covalent isomers and asymmetric electrostatically bound OCS$^{-} \cdot $ OCS structure behave differently in photoexcitation. 
The availability of low-lying excited states has been 
suggested to explain the autodetachment electron emission from the covalently bound isomers. 
The results in Fig. 5 show that the most pronounced low excited levels exist for the  C$_2$ symmetric isomers. The low-lying excited states do not appear in the asymmetric (OCS)$_2^{-}$ structure. Therefore, 
this calculations shows that the C$_2$ covalently bound
isomers are the most suitable for the autodetachment mechanism of electron emission.

In summary, three electronic isomers of the carbonyl sulfide dimer
anion are stabilized and investigated using the pseudopotential
based DFT computational methods. These three isomers of
(OCS)$_2^{-}$ have different geometrical and electronic properties,
and therefore their participation in various chemical and physical
processes is different. In particular, the Hirshfeld method
calculations show that a distribution of a negative charge strongly
depends on the structure and that in an asymmetric
isomer the charge is distributed as $-0.6 : -0.4$, where the
more bended moiety takes more charge.
The low-lying excited levels of (OCS)$_2^{-}$ isomers are calculated within the time-dependent density
functional method and the lowest excitations are found for the C$_2$ symmetric structure.
This work describes three possible structures of the carbonyl sulfide dimer anion.
These results are in rather good agreement
with the previous computational results obtained by different method \cite{Lineberger,Sanov}. 
Two similar covalent isomers are obtained, but electrostatically bound OCS$^{-} \cdot $ OCS structure
is not found in the present study. Instead weak covalent
OCS$^{-0.6} \cdot$ OCS$^{-0.4}$ asymmetric structure is minimized.
Many isomers, stable and metastable ones, may form under experimental conditions.
Therefore, the existence of all four isomers are possible in experiments, and all of them may contribute to
the time-resolved dynamics and solvation processes in $(OCS)_n^-$ clusters.
It is possible to assess the reliability of these calculations by
comparison with the results of the Gaussian code \cite{Lineberger,Sanov}. The two
lowest energy covalent isomers of the same symmetry are found using
both methods and their computationally optimized structural
parameters are close; the largest differences are below $2 \%$.
This agreement shows the precision and the reliability of both
techniques, as well as the strengths of all modern computational
methods in the studies of materials.
The details of the structure
and charge distribution in isomers of (OCS)$_2^{-}$ are also
important on a more general level because they play a role as a
prototype for the (OCS)$_n^{-}$ cluster ions and corresponding
solvation effects.
Because of the computational efficiency of the pseudopotential density functional theory method, it is possible to extend these calculations to larger anion clusters,
such as $(OCS)_n^-$ already studied in experiments described in Refs. 
\cite{Lineberger,Surber,Sanov}.
The pseudopotential time-dependent density functional techniques, used in this work to compare distributions od low-lying excited energy levels,
are now under intensive development. These methods
should be useful in an interpretation of experimental results on
the time-resolved dynamics of cluster anions and other nanostructures.

\begin{ack}
This work has been supported by the HR-MZOS projects
``Dynamical Properties and Spectroscopy of Surfaces
and Nanostructures'' and ``Electronic Properties of Surfaces and Nanostructures''.
The visualizations were done using the XCrySDen package \cite{Tone}.
I would like to thank the University Computing Center
SRCE  for their support and computer time.
\end{ack}

\end{document}